\documentclass[sigconf,authorversion]{acmart}

\AtBeginDocument{%
  \providecommand\BibTeX{{%
    \normalfont B\kern-0.5em{\scshape i\kern-0.25em b}\kern-0.8em\TeX}}}

\usepackage{bm}
\usepackage{bbm}

\copyrightyear{2020}
\acmYear{2020}
\setcopyright{acmlicensed}\acmConference[CIKM '20]{Proceedings of the 29th ACM International Conference on Information and Knowledge Management}{October 19--23, 2020}{Virtual Event, Ireland}
\acmBooktitle{Proceedings of the 29th ACM International Conference on Information and Knowledge Management (CIKM '20), October 19--23, 2020, Virtual Event, Ireland}
\acmPrice{15.00}
\acmDOI{10.1145/3340531.3412102}
\acmISBN{978-1-4503-6859-9/20/10}

\begin{document}

\title{Embedding Node Structural Role Identity into Hyperbolic Space}

\author{Lili Wang}
\affiliation{%
  \institution{Dartmouth College}
  \city{Hanover}
  \state{New Hampshire}
  \country{USA}}
\email{lili.wang.gr@dartmouth.edu}

\author{Ying Lu}
\affiliation{%
  \institution{Stony Brook University}
  \city{Stony Brook}
  \state{New York}
  \country{USA}}
\email{yinglu1@cs.stonybrook.edu}

\author{Chenghan Huang}
\affiliation{%
  \institution{Jefferies Financial Group LLC}
  \city{New York}
  \state{New York}
  \country{USA}}
\email{njhuangchenghan@gmail.com}

\author{Soroush Vosoughi}
\affiliation{%
  \institution{Dartmouth College}
  \city{Hanover}
  \state{New Hampshire}
  \country{USA}}
\email{soroush.vosoughi@dartmouth.edu}


\renewcommand{\shortauthors}{Wang et al.}

\begin{abstract}

Recently, there has been an interest in embedding networks in hyperbolic space, since hyperbolic space has been shown to work well in capturing graph/network structure as it can naturally reflect some properties of complex networks. However, the work on network embedding in hyperbolic space has been focused on microscopic node embedding. In this work, we are the first to present a framework to embed the structural roles of nodes into hyperbolic space. Our framework extends struct2vec, a well-known structural role preserving embedding method, by moving it to a hyperboloid model. We evaluated our method on four real-world and one synthetic network. Our results show that hyperbolic space is more effective than euclidean space in learning latent representations for the structural role of nodes.

\end{abstract}


\maketitle

\section{Introduction}

Most network embedding methods focus on preserving local structure information among connected vertices in their neighborhoods, like first-order, second-order, and high-order proximity. Using language models to preserve the microscopic structure of networks was first proposed
by Perozzi et al. in their work \texttt{DeepWalk} \cite{deepwalk}. This method uses random walks to generate random sequences of nodes from the network, which are then treated as sentences by a \emph{Skip-Gram} model \cite{word2vec}. 
Grover et al. \cite{node2vec} demonstrated that \texttt{DeepWalk} can not accurately capture the diversity of connectivity patterns in a network and introduced \texttt{node2vec}. They defined a flexible notion of a node's network neighborhood and designed a second-order random walk strategy to sample the neighborhood nodes. The method can smoothly interpolate between breadth-first sampling (BFS) and depth-first sampling (DFS). However, a limitation of these methods is that they can not capture structural role proximities. 

The structural role proximity depicts similarity between vertices serving similar ``roles'' in the network, such as being the center of a community, or a bridge between two communities. Different from the $k$th-order proximity, which captures the local similarity between nodes, the structural role proximity tries to discover the similarity between nodes far away from each other (or even disconnected) but sharing the equivalent structural roles. One of the early unsupervised methods for learning structural node embeddings is \texttt{RolX} \cite{rolx}. The method is based on enumerating various structural features for nodes in a network, finding the more suited basis vector for this joint feature space, and then assigning for every node a distribution over the identified roles. \texttt{struc2vec} \cite{struc2vec} determines the structural similarity between each node pair in the graph considering $k$-hop count neighborhood sizes. It constructs a weighted multilayer graph to generate a context for each node. \texttt{GraphWave} \cite{graphwave} uses one matrix factorization method based on the assumption that if two nodes in the network share similar structural roles, the graph wavelets starting at them will diffuse similarly across their neighbors. 

There also has been a relatively recent push for embedding networks into hyperbolic space. This has come with the realization that complex networks may have underlying hyperbolic geometry. This is because hyperbolic geometry can naturally reflect some properties of complex networks (such as the hierarchical and scale-free structures) \cite{krioukov2010hyperbolic}. An emerging network embedding approach is to embed networks into hyperbolic space \cite{nickel2017poincare,alanis2016manifold,de2018representation,muscoloni2017machine,HEAT,HHNE}. 
For instance, \texttt{HEAT} \cite{HEAT} learns embeddings form attributed networks and 
\texttt{HHNE} \cite{HHNE} learns embeddings form heterogeneous information network in hyperbolic space. 

However, to the best of our knowledge, none of the existing hyperbolic embedding methods can capture the structure role equivalence. To fill this gap, we present a framework to embed the structural roles of nodes into hyperbolic space. Our framework extends struct2vec, a well-known structural role preserving embedding method, by moving it to a hyperboloid model.

\section{Our Framework}
Let $G=(V, E)$ be a undirected and unweighted network, $V$ is a set of vertices and $E \subseteq V \times V $
is the set of unweighted edges between vertices in $V$. We consider the problem of representing a graph $ G=\left(V, E \right)$ as set of low-dimensional vectors into the $n$-dimensional hyperboloid $\left\{\mathbf{x}_{v} \in \mathbb{H}^{n} | v \in V\right\},$ with $n<<|V|.$ The described problem is unsupervised. Our framework consists of two parts: building the multi-layer graph which measures the structural similarity between node pairs, and using the context of each node generated by a biased random walk to learn hyperboloid embeddings.

\subsection{Constructing the Multi-layer graph}
The architecture presented in this paper can use any of the known approaches for node structural embeddings to generate the node context. In this paper, we extended \texttt{struct2vec}, the framework proposed by Ribeiro et al. \cite{struc2vec}. Let $Hop_{k}(u)$ denote the ordered sequence of the degree of the nodes at distance exactly $k$ from $u$ in $G$ (hop count).
The structural role similarity of two nodes $u$ and $v$ considering the set of nodes of distance $k$ from them can be defined as the similarity of the two ordered sequences $Hop_{k}(u)$ and $Hop_{k}(v)$. Note that these two sequences may not have equal sizes and their elements are integers in the range $[0, |G|-1]$. We use \emph{Fast Dynamic Time Warping} (FastDTW)\cite{fastdtw} to
measure the distance between two ordered degree sequences. The dynamic time warping algorithm (DTW) is able to find the optimal alignment between two arbitrary length time series, but has a quadratic time and space complexity that limits its use to only small time series data sets. The FastDTW is an approximation of DTW which limits both the time and space complexity to $O(n)$. Since elements of the sequences $Hop_{k}(u)$ and $Hop_{k}(v)$ are degrees of nodes, we adopt the following distance function of $i$th and $j$th element in the above two sequences for FastDTW as follows:
\begin{equation}
    DTW_{dis}(Hop_{k}^i(u), Hop_{k}^j(v)) =\frac{\max (Hop_{k}^i(u), Hop_{k}^j(v))}{\min (Hop_{k}^i(u), Hop_{k}^j(v))}-1 
    \label{distance}
\end{equation}
Instead of measuring the absolute difference of degrees, this distance measures the relative difference which is more suitable for degree differences. 
The structural role distance of two nodes $u$ and $v$ considering their $k$-hop neighborhoods can be defined as:
\begin{equation}
    distance_{k}(u,v)=\sum_{i=0}^k{DTW(Hop_{i}(u),Hop_{i}(v))}
\end{equation}

Next, we construct a multilayer weighted graph $M$ that encodes the structural similarity between nodes. Each layer $k=0,1...,diameter$ is constructed by a weighted undirected complete graph with all the nodes of the original graph $G$. The edges of $M$ inside layer $k$ are defined as:
\begin{equation}
    w(M_{k}^u, M_{k}^v)=e^{-distance_{k}(u, v)}, \quad k=0, \dots, diameter
\end{equation}
Note that if a node $u$ has too many or too few structurally similar nodes in the current layer $M_{k}$, then it should change
layers to obtain a more refined context. By moving up
one layer the number of similar nodes will decrease, and by moving down one layer the number of similar nodes will increase. Thus, we define the inter-layer edges as follows:
\begin{equation}
\begin{split}
    &w(M_{k}^u,M_{k+1}^u)=\log \left(L_{k}(u)+e\right), \quad k=0, \dots, diameter-1\\
     &w(M_{k}^u,M_{k-1}^u)=\text { 1, } , \quad k=1, \dots, diameter
\end{split}
\end{equation}
where $L_{k}(u)$ denotes how many nodes are structurally similar in layer $k$, which is the number of incoming edges to $u$ that have weight larger than the average weight of layer $k$, more specifically:
\begin{equation}
    L_{k}(u)=\sum_{v \in V} \mathbbm{1}\left(w(M_{k}^u, M_{k}^v)>\frac{\sum\limits_{u',v' \in V}w(M_{k}^{u'}, M_{k}^{v'})}{\tbinom{|V|}{2}}\right)
\end{equation}
We then adopt a random walk method to obtain the structural preserving context of each node. For each step, it can either walk inside one layer or walk between layers. We define the layer-change constant $\alpha$, such that for each step the probability of staying in the current layer is $1-\alpha$ and the probability of going up or down one layer is $\alpha$. Thus, given the current node $M_{k}^u$, the normalized probability of moving to a current layer node $M_{k}^v$ is:
\begin{equation}
    p(M_{k}^v|M_{k}^u)=(1-\alpha) \frac{ w(M_{k}^u, M_{k}^v)}{\sum\limits_{u,v \in V}w(M_{k}^u, M_{k}^v)}
\end{equation}
The normalized probability of moving to a node in the layer above, $M_{k+1}^u$, is:
\begin{equation}
    p(M_{k+1}^u|M_{k}^u)= \alpha\frac{ w(M_{k}^u, M_{k+1}^u)}{w(M_{k}^u, M_{k+1}^u)+w(M_{k}^u, M_{k-1}^u)}
\end{equation}
And similarly, the normalized probability of moving to a node in the layer below, $M_{k-1}^u$, is:
\begin{equation}
    p(M_{k-1}^u|M_{k}^u)= \alpha\frac{ w(M_{k}^u, M_{k-1}^u)}{w(M_{k}^u, M_{k+1}^u)+w(M_{k}^u, M_{k-1}^u)}
\end{equation}

\subsection{Learning a Hyperboloid Model}
Finally, we train a \emph{hyperboloid} model on the generated random walk sequences to obtain structural role preserving embeddings. Hyperbolic space is a homogeneous space with constant negative curvature. It can not be embedded into the Euclidean space without distortion, however, there are several hyperbolic models that allow calculation of gradients. The most commonly used ones are \emph{hyperboloid}, \emph{Poincar\'e ball}, and \emph{Poincar\'e half-space}. Unlike previous works using Poincar\'e ball model and approximate gradients, we use the hyperboloid model for network embedding because the gradient computation of this model is exact \cite{gradient} and we can adopt a Support Vector Machine (SVM) on it \cite{svm}.
\subsubsection{Review of the Hyperboloid Model}
The hyperboloid model has many similarities to the sphere model. Analogous to the sphere in the ambient Euclidean space, the hyperboloid model can be viewed as a ``pseudo-sphere'' in an ambient space called the $Minkowski\,space$.
Consider an (n+1)-dimensional space equipped with an inner product whose form is given by:
\begin{equation}
<u,v>_{M} = \sum_{i=1}^{n}{u_iv_i}\,-\,u_{n+1}v_{n+1}.
\label{eq1}
\end{equation}
We use $\mathbb{R}^{n:1}$ for the notation of Minkowski space.
Analogous to the unit sphere in Euclidean space, the hyperboloid can be described using the following equation:
\begin{equation}
\mathbb{H}^n=\left\{x\in \mathbb{R}^{n:1}\,|\,<x,x>_M = -1 \right\},
\end{equation}
For a given vector $p\in \mathbb{H}^n$, the tangent space at that point is a set of points with the form
\begin{equation}
T_p\mathbb{H}^n = \left\{x\in\mathbb{R}^{n:1}\,|\,<p,x>_M = 0 \right\}.
\end{equation}

\subsubsection{Gradient calculation on the Hyperboloid $\mathbb{H}^n$}
Analogous to the case of sphere, The calculation of the gradient of a given function $E$ defined on $\mathbb{H}^n$ has several steps \cite{gradient}.

$Step\,1:$ Calculate the gradient of E in the ambient space, i.e. 
\begin{equation}
    \nabla_p^{R^{n:1}}{E}=\left(\frac{\partial{E}}{\partial{x_1}}\vert_p\,,...,\,\frac{\partial{E}}{\partial{x_n}}\vert_p\,,\,-\frac{\partial{E}}{\partial{x_{n+1}}}\vert_p\right)\in\mathbb{R}^{n:1}.
    \label{eq9}
\end{equation}

$Step\,2:$ Project that vector onto the tangent space $T_p\mathbb{H}^n$. Notice that the sign is flipped in the expression of the projected vector: 
\begin{equation}
\nabla_p^{\mathbb{H}^n}{E}=\nabla_p^{R^{n:1}}{E}+<p,\nabla_p^{R^{n:1}}{E}>_M\cdot{p}\in{T_p}\mathbb{H}^n.
\label{eq10}
\end{equation}

$Step\,3:$ Map the gradient vector $\nabla_p^{\mathbb{H}^n}{E}$ onto the hyperboloid. This operation is called \emph{exponential map}.
\begin{equation}
\rm {Exp_p(v)}=\rm{cosh}(\|v\|)p+\rm{sinh}(\|v\|)\frac{v}{\|v\|}\in\mathbb{H}^n,
\label{eq11}
\end{equation}
where $v=\nabla_p^{\mathbb{H}^n}{E}\in{T_p}\mathbb{H}^n$.

\subsubsection{Hyperboloid Embedding Learning}
After generating the random walk sequences, we use a sliding window to scan all the sequences and add pairs of nodes that appear within the window to a multi-set $O$ as all the positive sample pairs. Note that different from common sampling methods, each pair of nodes $u$ and $v$ can appear multiple times in $O$. Intuitively, the number of times a pair is sampled indicates the importance of that pair. In prior work \cite{nickel2017poincare}, Nickel et al used the distance of two nodes to define the possibility of a link. Similarly, we define the structural role similarity of two nodes to be their distance in the embedded hyperbolic space: nodes close to each other share a high similarity and vice versa. We define the structural role distance between nodes $u$ and $v$ as
\begin{equation}
\frac{\exp \left(-d_{\mathrm{H}^{n}}^{2}\left(\mathbf{x}_{u}, \mathbf{x}_{v}\right) \right)}{\sum_{v^{\prime} \in V} \exp \left(-d_{\mathrm{H}^{n}}^{2}\left(\mathbf{x}_{u}, \mathbf{x}_{v^{\prime}}\right) \right)},
\label{eq2}
\end{equation}
where $\mathbf{x}_{u}$ is the embedding of node $u$ in the hyperboloid model. $d_{\mathrm{H}^{n}}^{2}\left(\mathbf{x}_{u}, \mathbf{x}_{v}\right)$ can be calculated by $d_{\mathrm{H}^{n}}\left(\mathbf{x}_{u}, \mathbf{x}_{v}\right) = arccosh(-<u,v>_M)$. 
However, computing the gradient of Equation \ref{eq2} involves a summation over all the nodes of $V$ and is inefficient for large networks. To address this, we leverage the negative sampling method which samples a small number of negative objects to enhance the influence of positive samples. As a result, our loss function $L$ for an embedding $\Theta=\left\{\mathbf{x}_{u} \in \mathbb{H}^{n} | u \in V\right\}$ can be written as following:
\begin{equation}
\begin{split}
& L(\Theta)=-\frac{1}{|O|} \sum_{(u, v) \in O} \log \\ &\frac{\exp \left(-d_{\mathrm{H}^{n}}^{2}\left(\mathbf{x}_{u}, \mathbf{x}_{v}\right)  \right)}{\exp \left(-d_{\mathrm{H}^{n}}^{2}\left(\mathbf{x}_{u}, \mathbf{x}_{v}\right)  \right)+\sum\limits_{i<=M, v^{\prime}_{i} \in P(u)} \exp \left(-d_{\mathrm{H}^{n}}^{2}\left(\mathbf{x}_{u}, \mathbf{x}_{{v^{\prime}}_{i}}\right) \right)}
\end{split}
\end{equation}
where $P(u):=\{v |(u, v) \notin O, v \in V\}$ is the negative sampling set with probability proportional to the occurrence frequency of $v$ in $O$, $M$ is the number of negative samples. The calculation of its gradient follows Equation. \ref{eq9}, \ref{eq10} and \ref{eq11}, which enables the gradient decent for model learning.

\section{Experiments}
We use the same five datasets used by Leonardo et al \cite{struc2vec}: one synthetic barbell graph; four real-word datasets: Brazilian, American and European air-traffic network and karate network. 
\subsection{Model Training}\label{sec:para}
For random on the multi-layer graph,  the layer-change constant $\alpha$ is set to 0.7, and we do 8 random walks from each node in the training set with the length of 10. (Contrast that with the classic struct2vec in Euclidean space that needs to set the number of random walks to 80. Our method reduces the need for random walks, which are computationally expensive, by 90\%) For training the hyperboloid model, we use a sliding window of size 3 to generate positive samples. For the hyperboloid embedding learning, we generate 20 negative samples for each positive ones, and use the learning rate of 1 and a batch size of 50 to train 5 epochs.

\subsection{Barbell Graph}

\begin{figure}[ht]%
\includegraphics[width=\linewidth]{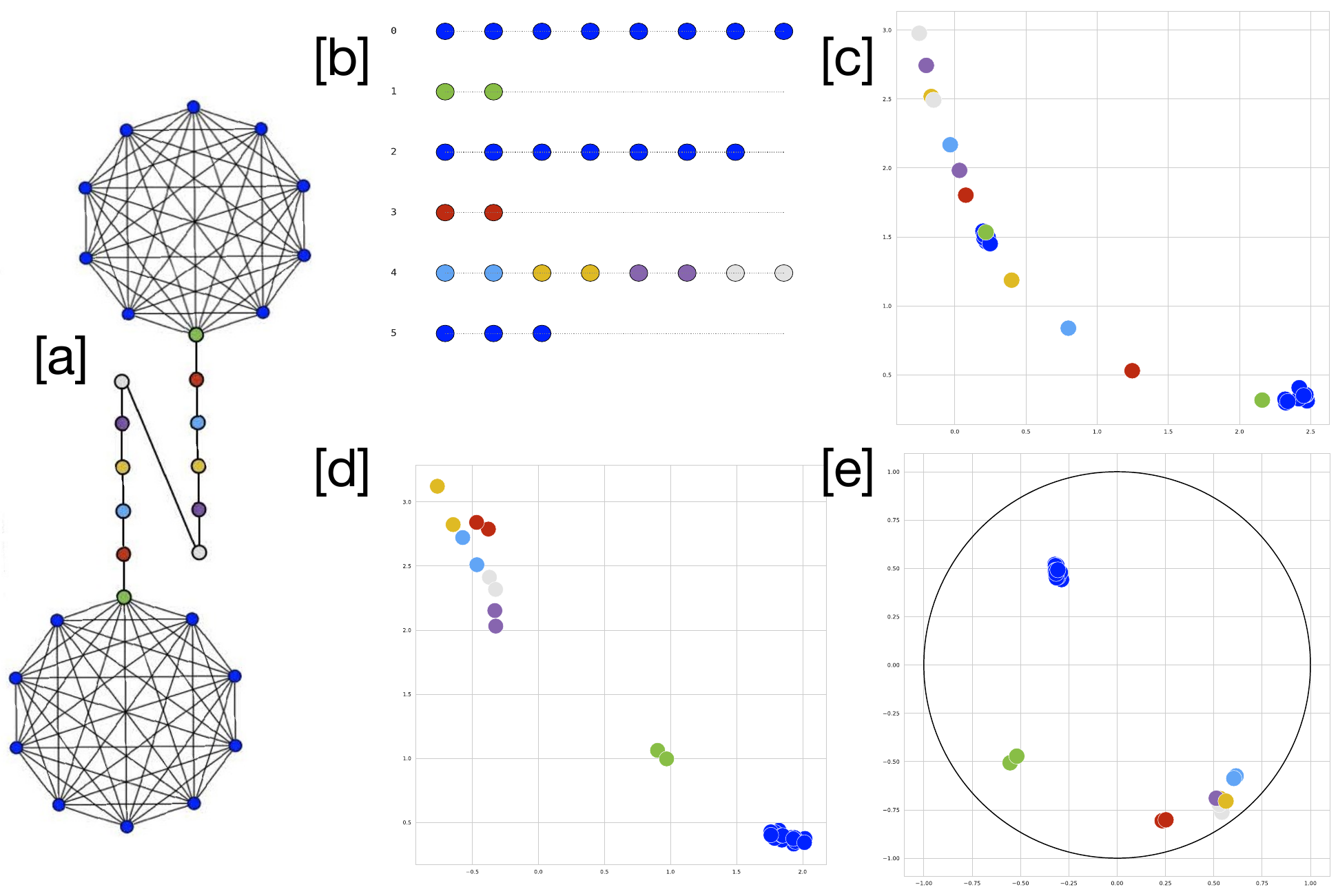}
\caption{(a) the Barbell graph\cite{struc2vec} used in our experiment (b) Roles identified by \texttt{RolX} (c) node2vec (d) struct2vec (e) hyperboloid (our method) visualization results of embeddings on the barbell graph. }
\label{barbell}
\end{figure}

\begin{figure}[ht]%
\includegraphics[width=\linewidth]{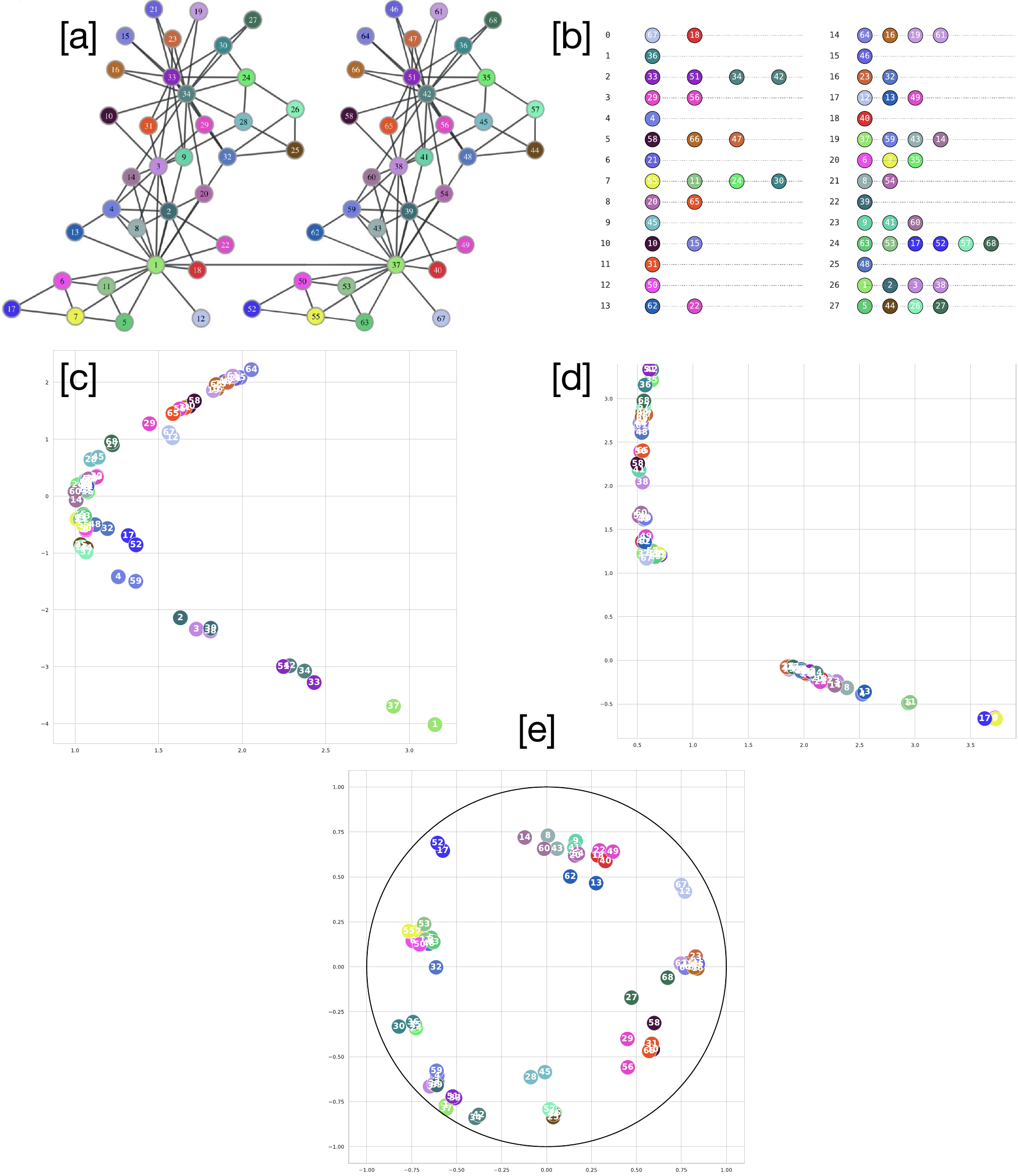}
\caption{(a) the karate network \cite{struc2vec} used in our experiment (b) Roles identified by \texttt{RolX} (c) node2vec (d) struct2vec (e) hyperboloid (our method) visualization results of embeddings on the karate network.}
\label{karate}
\end{figure}

We consider the barbell graph which consists of two complete subgraphs connected by a long path. Figure \ref{barbell}(a) shows the barbell graph used in the experiment, where the structurally equivalent nodes
have the same color. The result of \texttt{RolX} is in Figure \ref{barbell}(b), although \texttt{RolX} captures some structural role identity, all the blue nodes are placed in three different roles (0,2 and 5). Also, role 4 contains all the nodes in the path, but actually they are not exactly similar. Figure \ref{barbell}(c) shows the results of \texttt{node2vec}, it does not capture structural role identities and the nodes of two parts of the complete graph are placed separately, along with the nodes in the path close to them. Figure \ref{barbell}(e) shows our results on a 2-dimensional Poincar\'e ball, compared with \texttt{struct2vec} results in Euclidean space (Figure \ref{barbell}(d)), our method captures structural equivalence more accurately. Moreover, we only do 8 random walks of length 10 from each node in the hyperboloid model, and \texttt{struct2vec} needs to set the number of random walk to 80 to generate an accurate result, which also indicates the superiority of hyperbolic space in learning structural role equivalence. 

\subsection{Karate Network}
The Zachary's Karate Club \cite{karate} is a network of 34 nodes: each node represents a club member and edges among them denote if two members have interacted. The network used in the experiment (Figure \ref{karate}(a) ) is composed of two copies of the Karate Club network, where each node has a mirror node and one edge has been added between mirrored node pairs 1 and 37. Figure \ref{karate}(b) shows the roles identified by \texttt{RolX}, only 7 out of 34 corresponding pairs are placed in the same role. Result of \texttt{node2vec} is shown in Figure \ref{karate}(c), since this method only captures microscopic structural information, the two parts of the network are placed separately since there is only one edge that connects them. The corresponding pairs of our result on a 2-dimensional Poincar\'e ball (Figure \ref{karate}(e)) are more close than the result of \texttt{struct2vec} (Figure \ref{karate}(d)). Moreover, different roles' embeddings generated by \texttt{struct2vec} are more likely to bunch together in Euclidean space. In hyperbolic space, however, these embeddings are located more sparsely, which indicates a better ability to distinguish different roles.

\subsection{Node classification}

We also test our method on three real-world datasets provided by Leonardo et al \cite{struc2vec}: Brazilian, American and European air-traffic networks. The nodes
correspond to airports and edges indicate the existence of commercial fights. For each airport, one of four possible labels is assigned corresponding to their activity (divided evenly into four quartiles). Thus, each class represents a "role" played by the airport (e.g, major hubs). The task here is to predict the role of an airport. We train all the three models on each network to get embeddings and use a 10-fold cross-validation for the evaluation.  For our model, we use a hyperbolic SVM \cite{svm} as the classifier, and for the other two Euclidean models \texttt{struct2vec} and \texttt{node2vec}, we use the classic Euclidean SVM. Table \ref{nodeclassification} shows the node classification results where our model outperforms the baselines.

\begin{table} 
\centering
  \caption{Micro F1 score of our model (Hyperboloid) vs all the baselines for the node classification task}
   \label{nodeclassification}
\begin{tabular}{|l|l|l|l|}
\hline
     & Brazilian & American & European  \\
     \hline
       Hyperboloid  & \textbf{0.780} & \textbf{0.670} &\textbf{0.581}  \\
       Struct2vec &  0.732 & 0.651 & 0.577   \\
       Node2vec &  0.267 & 0.473 & 0.329  \\ 
     
\hline
\end{tabular}

\end{table}

\section{Conclusion}
In this paper, we present a novel method for embedding nodes of a network into hyperbolic space which preserves structure role information. To the best of our knowledge, this is the first attempt at a hyperbolic model that can learn node structural role proximity. Our algorithm outperforms several baselines on a synthetic barbell graph and four real-world temporal datasets for embeddings visualization and node classification. The code and data for this paper will be made available upon request. 

\bibliographystyle{ACM-Reference-Format}
\bibliography{sample-base}

\end{document}